\documentclass[a4paper,11pt]{article}
\usepackage{amsmath}
\usepackage{graphicx}
\usepackage{amssymb}
\usepackage{amsfonts}
\usepackage{latexsym}

\def\beq{\begin{eqnarray}}
\def\eeq{\end{eqnarray}}
\def\ln{\,\mbox{ln}\,}

\def\tr{\,\mbox{tr}\,}
\def\Tr{\,\mbox{Tr}\,}

\def\al{\alpha}
\def\be{\beta}

\def\ga{\gamma}\def\de{\delta}
\def\vp{\varepsilon}

\def\ze{\zeta}

\def\ka{\kappa}
\def\la{\lambda}
\def\na{\nabla}
\def\pa{\partial}

\def\rh{\rho}
\def\si{\sigma}
\def\om{\omega}
\def\ph{\varphi}
\def\ta{\tau}

\def\Ga{\Gamma}

\def\La{\Lambda}

\newcommand{\eq}[1]{(\ref{#1})}
\newcommand{\n}[1]{\label{#1}}
\newcommand{\nn}{\nonumber}
\usepackage{indentfirst}
\usepackage{mathrsfs} 

\usepackage{cite}
\usepackage{color}

\usepackage{soul}

\begin{document}

\renewcommand*{\thefootnote}{\fnsymbol{footnote}}

\begin{center}
{\large\sc
Vilkovisky unique effective action in quantum gravity}

\vskip 6mm
{\small \bf Breno L. Giacchini,}$^a$%
 \footnote{E-mail address: breno@sustech.edu.cn}
\quad {\small \bf Tib\'erio de Paula Netto}$^a$%
 \footnote{E-mail address: tiberio@sustech.edu.cn}
\quad and \quad {\small \bf Ilya L. Shapiro}$^{b}$%
\footnote{On leave from Tomsk State Pedagogical University.
\ E-mail address: ilyashapiro2003@ufjf.br}
\vskip 6mm

(a) Department of Physics,
\ Southern University of Science and Technology,
\\ Shenzhen, \ 518055, \ China
\vskip 2mm

(b) Departamento de F\'{\i}sica, \ ICE, \
Universidade Federal de Juiz de Fora,
\\ Juiz de Fora, \ 36036-900, \ MG, \ Brazil
\vskip 2mm


%
\end{center}
\vskip 4mm

\centerline{\uppercase{abstract}}
\vskip 2mm
\begin{quotation}

\noindent
The divergent part of the one-loop Vilkovisky unique effective action for quantum
Einstein gravity is evaluated in the general parametrization of the quantum field,
including the separated conformal factor.
The output of this calculation explicitly
demonstrates the parametrization and conformal gauge independence of the unique 
effective action with the configuration space metric chosen following Vilkovisky's 
prescription.
\vskip 4mm

\end{quotation}

\renewcommand*{\thefootnote}{\arabic{footnote}}
\setcounter{footnote}{0}

\section{Introduction}
\label{Sec0}

The off-shell effective action in gauge theories depends on the
choice of the gauge fixing and the parametrization of quantum fields.
On the other hand, in the modified versions of effective
action proposed by Vilkovisky
\cite{Vil-unicEA} and DeWitt \cite{DeWitt-ea} there is  no gauge or
parametrization ambiguity. The purpose of the present
work is to evaluate the divergent part of the one-loop Vilkovisky
effective action for the quantum version of Einstein gravity in a general
parametrization of the quantum field and explicitly verify the
independence of this construction on the parametrization.

The classical action of the theory of our interest has the form
\begin{equation}
\label{action}
S (g_{\mu\nu})
\,=\, - \frac{1}{\kappa^2} \int \text{d}^D x \sqrt{\vert g \vert}
\big(R + 2 \La\big),
\end{equation}
where $\,G = \ka^2/(16 \pi)$ is the ($D$-dimensional) Newton
constant and $\La$ is the
cosmological constant. There is an extensive literature on the
derivation and analysis of one-loop and two-loop divergences in the
theory
\eqref{action}. The first calculations were performed in \cite{hove}
for gravity coupled with the minimal scalar field and in \cite{dene}
for gravity coupled to an electromagnetic field. The calculation in the
nonminimal gauge was pioneered in \cite{KTT}. The parametrization
dependence was explored in
\cite{FirPei,Kalmykov:1995fd,Kalmykov:1998cv} and, in a more
general form, in the more recent Ref.~\cite{JDG-QG}. In what follows
we shall use some technical developments of the latter work, which
can be also consulted for further references.

The unique effective action of Vilkovisky is independent of the
parametrization of quantum fields by construction. On the other hand,
this construction becomes complicated in gauge theories, where one
has to combine corrections compensating gauge and
parametrization ambiguities. In this regard, a special case is the
two-dimensional quantum gravity. It was noted in \cite{Banin-PLB}
that, in this particular example, the gauge and parametrization
ambiguities mix in such a way that the unique effective action
turns out to depend on the gauge fixing. The origin of this
contradictory result is that the unique effective action depends
on the choice of the metric in the configuration space,
or the space of the quantum fields, in the background field
formalism, as it was anticipated in the very first work~\cite{Vil-unicEA}.
In gravity, the configuration-space metric has one
arbitrary parameter $a$, and it happens that in the $D=2$ covariant 
formulation of the metric-scalar theory (see, \textit{e.g.}, \cite{OdSh90} 
and \cite{GKV} for the review) this parameter depends on the gauge 
fixing, because of the reduced number of the physical degrees of 
freedom~\cite{Banin-PLB}.
As a result, the metric in the configuration space depends on the 
gauge-fixing parameters even if the bilinear form of the action acquires 
the simplest minimal form. The four-dimensional quantum gravity in the 
conformal parametrization has a seeming similarity with the mentioned 
$D=2$ case, because the metric in the configuration space also depends 
on a gauge parameter, namely, the conformal gauge-fixing parameter 
$\la$~\cite{FirPei,JDG-QG}. Thus, one could suspect that some gauge or
parametrization dependence in the off-shell unique effective action 
may persist in this case too.

Let us note that the Vilkovisky-DeWitt approach in 
quantum gravity opens the way for formulating the \textit{exact} renormalization group 
flow for the cosmological and Newton constants and for the full set of
higher-derivative terms which should be added to the
Einstein-Hilbert action in the framework of effective field theory
\cite{UEA-RG} (see also~\cite{TV90,BarberoG.:1993cw} for
previous analysis of the renormalization group based on the unique
effective action in quantum gravity).  This makes the verification of
the consistency of this approach in $D=4$ even more interesting.

The outline of the paper is as follows. Section~\ref{Sec1} briefly
reviews the formalism of Vilkovisky's effective action. The main
objective of this section is to make the paper self-consistent and to
fix the  notations.
In Sec.~\ref{Sec2}, we formulate the one-loop quantum gravity using
the background field method in a general non-conformal parametrization
of quantum field and a special minimal gauge.
The metric in the space of the fields, the Christoffel symbols,
and the improved bilinear form of the classical action
are derived in Sec.~\ref{Sec3}. It is shown that the coefficients
related to the parametrization nonlinearity are compensated by this
correction.
The corresponding one-loop divergences of the Vilkovisky effective
action are computed, in the minimal DeWitt gauge, in Sec.~\ref{Sec4}.
In Sec.~\ref{ApB}, the result is generalized to the most general,
conformal parametrization of the quantum metric.
Finally, in Sec.~\ref{Sec5}, we draw our conclusions.

In this paper we adopt the condensed notations of
Refs.~\cite{BDW-65} and \cite{Bavi83-85}.

\section{Vilkovisky effective action: A short review}
\label{Sec1}

Vilkovisky's proposal for defining a parametrization-independent
effective action \cite{Vil-unicEA}
is based on the following observation: even though
the classical action $S(\ph)$ is a scalar in the space $\mathscr{M}$
of fields $\ph^i$, the generating functional of vertex functions
(effective action) is not a scalar functional of the corresponding
mean fields. In the simplest, one-loop approximation the effective
action depends on the  Hessian of the action,
$S_{,ij}=\frac{\de^2 S}{\de \ph^i \de \ph^j}$,
which does not transform as a tensor under field redefinitions
$\ph^i = \ph^i(\ph'^j)$.

To provide the scalar nature of the effective action, in
Ref.~\cite{Vil-unicEA} it was introduced an affine structure
compatible with the metric $G_{ij}$ in the space $\mathscr{M}$.
For given two close points $\ph^i$ and $\ph^{\prime i}$, there
exists a unique geodesic curve $x^i(\la) \subset \mathscr{M}$ with
affine parameter $\la \in [0,1]$ connecting them,  $x^i(0) = \ph^i$
and $x^i(1) = \ph^{\prime i}$. Then, defining the two-point quantity
$\si^i(\ph^\prime,\ph) = \frac{dx^i (\la)}{d\la}\big|_{\la=1}$ (the
tangent vector to the geodesic at $\ph^{\prime i}$, see
{\it e.g.}~\cite{BDW-65,J.L.Synge:1960zz}), the modified definition
of the effective action has the form
\beq
\exp i \Ga(\ph)
\,=\,
\int \mathcal{D} \ph' \mu(\ph^{\prime})
\,\exp\left\lbrace
i \left[ S(\ph^{\prime}) + \si^i(\ph,\ph^{\prime}) \Ga_{,i}(\ph) \right]
\right\rbrace,
\label{UEA_def}
\eeq
where $\mu(\ph^{\prime})$ is an invariant functional measure and
the comma denotes functional differentiation with respect to $\ph^i$.
The effective action $\Ga (\ph)$ constructed in this way is a scalar
under field reparametrizations because $\si^i(\ph,\ph^\prime)$ behaves
as a vector with respect to $\ph^{i}$ and as a scalar with regard to
$\ph^{\prime i}$.

A qualitatively similar construction can be done for gauge theories,
to restore the off-shell gauge independence, given that
the effective actions calculated in
different gauges are connected by changes of variables (in general,
in the form of a canonical
transformation~\cite{Voronov:1982ur,Voronov:1984kq,Voronov:1982ph}).
However, in this case, the prescription~\eqref{UEA_def} cannot be used
directly since it is necessary to factor out the gauge group
$\,\mathscr{G}\,$ in the functional integral. Namely, one has to take
into account the gauge orbits and define an affine connection in the
configuration space $\mathscr{M}/\mathscr{G}$ of physical fields.
For the sake of
simplicity, we assume that the generators $R^i_\al$ of gauge
transformations are linearly independent and their algebra is closed,
$R^i_{\be,j} R^j_\al - R^i_{\al,j} R^j_\be = F^\ga_{\al\be} R^i_\ga$,
with the structure functions $F^\ga_{\al\be}$ being independent of the
fields.
Let the classical action be invariant under gauge transformations
$\de \ph^i = R^i_\al \, \xi^\al$,
\beq
\vp_i\, R^i_\al  \,=\, 0,
\qquad
\vp_i \equiv S_{,i}.
\eeq
Given a metric $G_{ij}$ on $\mathscr{M}$ one can define
the projection operator on $\mathscr{M}/\mathscr{G}$
\cite{Vil-unicEA,Fradkin:1983nw}
\beq
\label{Projetor}
P^i_j \,=\, \de^i_j - R^i_\al N^{\al\be} R^{k}_\be G_{kj}
\,,
\eeq
where $N^{\al\be}$ is the inverse of the metric on $\mathscr{G}$,
\beq
N_{\al\be} \,=\, R^i_\al G_{ij} R^j_\be .
\label{N_def}
\eeq
Then the projected metric is
\begin{equation}
\begin{split}
{G}_{i\,j}^{\perp\perp} & \, \equiv \, P^k_i  G_{kl} P^l_j
\, = \, G_{ij} - G_{ik} R^k_\al N^{\al\be} R^l_\be G_{lj}.
\end{split}
\end{equation}
The affine connection $\,\mathscr{T}^k_{ij}\,$ on the physical
configuration space can then be obtained by requiring its
compatibility with the metric ${G}_{i\,j}^{\perp\perp}$ {\it i.e.}
$\na_k {G}_{i\,j}^{\perp\perp} = 0$ (see \textit{e.g.}
\cite{Kunstatter:1986qa,Huggins:1987zw}). This
yields~\cite{Vil-unicEA}
\beq
\mathscr{T}^k_{ij}
\,=\, \Ga^k_{ij} +  T^k_{ij},
\eeq
which consists of the Christoffel symbol $\,\Ga^k_{ij} \,$ calculated
with the metric $G_{ij}$,
\beq
\label{Gkij}
\Ga^k_{ij} \,=\, \frac12 \, G^{kl} ( G_{il,j} + G_{jl,i} - G_{ij,l} ),
\eeq
and a nonlocal part $\, T^k_{ij} \,$ related to the gauge constraints
on the connection,
\beq
\label{Tijk}
T^k_{ij} \,=\, - 2 G_{(i|l} R^l_\al N^{\al\be}
\mathscr{D}_{|j)} R^k_\be
+ G_{(i|l} R^l_\al N^{\al\be} R^m_\be (\mathscr{D}_m R^k_\ga)
N^{\ga\de} R^n_\de G_{n|j)}.
\eeq
The parentheses in the indices represent symmetrization in the pair
$(i,j)$, and $\mathscr{D}_i$ denotes the covariant derivative
calculated with the Christoffel connection $\Ga^k_{ij}\,$. The
nonlocality of \eq{Tijk} is due to the fact that $N_{\al\be}$ is a
differential operator and thus its inverse $N^{\al\be}$ is formally
a Green's function. In addition to that, this procedure provides the
measure $\mu(\ph)$ of the Faddeev-Popov quantization; see,
{\it e.g.}, Refs.~\cite{Ellicott:1989mi,bookPT}. The effective
action~\eqref{UEA_def} constructed using the geodesic distance
based on the connection $\,\mathscr{T}^k_{ij}\,$ is therefore
reparametrization invariant, gauge invariant and gauge independent.
For this reason, this object is often called
\textit{unique effective action}\footnote{Another gauge- and
parametrization-invariant effective action was proposed by
DeWitt~\cite{DeWitt-ea}. Since both definitions coincide at the one-loop level, we do
not present this construction. We remark, however, that for calculations in higher-loop orders it is necessary to use the Vilkovisky-DeWitt formalism, as the simplest form~\eqref{UEA_def} may generate nonlocal divergences~\cite{Rebhan:1986wp,Rebhan:1987cd} (see also~\cite{Ellicott:1987ir}).}.

Performing the loop expansion of the Vilkovisky effective action
\eq{UEA_def}, one gets
\beq
\Ga (\ph) = S( \ph) + \bar{\Ga}^{(1)} (\ph)
+ \bar{\Ga}^{(2)} (\ph) + \cdots,
\qquad\qquad
\mbox{$\hbar = 1$},
\eeq
where the one-loop quantum contribution is given by \cite{Vil-unicEA}
\beq
\label{EA1loop0}
\bar{\Ga}^{(1)}
\,=\,
\frac{i}{2} \Tr \ln G^{ik} (\mathscr{D}_k \mathscr{D}_j S
- T^l_{kj} \vp_l - \chi^\al_{,k} Y_{\al\be} \chi^{\be}_{,j} )
- i \Tr \ln M^\al_\be.
\eeq
As usual, in pure quantum gravity we can use $\ka$ as a loop
expansion parameter, instead of $\hbar$. Here, $\chi^\al$ is a
gauge condition introduced by the gauge-fixing action
\beq
S_{\text{GF}} = - \frac{1}{2} \chi^\al  Y_{\al\be} \chi^\be \, ,
\eeq
$Y_{\al\be}$ is a nondegenerate weight function (the
$\chi^\al$-space metric) and $M^\al_\be = \chi^\al_{,i} R^i_\be$
is the Faddeev-Popov ghost matrix. Comparing~\eqref{EA1loop0} to
the loop expansion of the standard effective action, one notes that the
second functional derivative of the classical action has been replaced
by the second covariant variational derivative.

From the technical side, the computation of \eq{EA1loop0} is, in
general, a very complicated task because of the nonlocalities of
the term $\, T^k_{ij} \,$. For this reason, most of the evaluations
found in the literature use some kind of DeWitt
gauge~\cite{DeWitt:1967ub}, for which
\beq
\n{de-gauge}
 \chi^\al_{,i}  \,=\, - Y^{\al\be} G_{ij} R^j_\be \,.
\eeq
The following observation is in order. It is quite common in the
literature (see \textit{e.g.}~\cite{Fradkin:1983nw,Huggins:1987zw,Rebhan:1986wp})
the use of the singular version of \eq{de-gauge}, $\chi^\al_{,i} = 0$,
also known as  Landau-DeWitt gauge. Such a gauge choice is convenient
as it yields ${T}^k_{ij} = 0$. Thus, in theories whose
field space $\mathscr{M}$ is flat, at one-loop level, the traditional
effective
action evaluated in the Landau-DeWitt gauge is equal to 
Vilkovisky's one~\cite{Fradkin:1983nw}. This gauge, however, is
not so auspicious in gravity theories because the geometry of
$\mathscr{M}$ is nontrivial~\cite{Fradkin:1983nw,Rebhan:1986wp,
Ellicott:1987ir}. A remarkable exception is the one-loop
divergences related to the cosmological constant and Einstein-Hilbert
term in quantum general relativity. In fact, it turns out that for
the Vilkovisky's choice of metric $G_{ij}$ in the space of fields the
$\Ga^k_{ij}$-correction in Eq.~\eq{EA1loop0} does not give any new
contribution to these terms; therefore, they can be directly obtained
by using the Landau-DeWitt gauge in the context of the usual
definition of the effective action~\cite{Fradkin:1983nw}.
As here we are interested in evaluating also the divergences
related to curvature-squared terms, for practical reasons we choose
to use the nonsingular version of the DeWitt gauge and deal with the
nonlocalities in the connection.

The purpose of the present work is to evaluate the divergent part of
\eq{EA1loop0} for the quantum gravity based on the general relativity.
In this calculation, we follow the reduction method introduced in
Ref.~\cite{Bavi83-85}, which mainly consists in making a power series
expansion in the equations of motion $\vp_i $ and applying the
generalized Schwinger-DeWitt technique. By using the DeWitt gauge
\eq{de-gauge} and the Ward identities, it is possible to
write~\eqref{EA1loop0} in the form~\cite{Bavi83-85}
\beq
\label{VEA}
\bar{\Ga}^{(1)}
= \frac{i}{2} \Tr \ln \hat{H} - i \Tr \ln \hat{N}
- \frac{i}{2} ( \Tr \hat{U}_1 - \Tr \hat{U}_2 )
- \frac{i}{4} \Tr \hat{U}_1^2 + O (\vp^3) ,
\eeq
where  $\hat{N} = Y^{\al\ga} N_{\ga\be}\,$ and $\,N_{\al\be}$ was
defined in \eqref{N_def},
\beq
\label{H}
\hat{H}
\,=\,
G^{ik} (\mathscr{D}_k \mathscr{D}_j S
- \chi^{\al}_{,k} Y_{\al\be} \chi^{\be}_{,j})
\eeq
takes into account the nontrivial geometry of the space of fields
$\mathscr{M}$,  and
\begin{align}
\label{U1op}
\hat{U}_1 &= N^{\al\ga} R^i_\ga (\mathscr{D}_i R^j_\de)
\vp_j N^{\de \si} Y_{\si \be} \, ,
\\
\label{U2t}
\hat{U}_2 &= N^{\al\ga} (\mathscr{D}_i R^k_\ga) \vp_k (H^{-1})^{ij}
(\mathscr{D}_j R^l_\de) \vp_l N^{\de\si} Y_{\si \be}
\,
\end{align}
are two nonlocal operators responsible for restoring the off-shell gauge
independence of the one-loop effective action. In \eq{U2t},
$\hat{H}^{-1}$ is defined by the relation
$\, \hat{H} \cdot \hat{H}^{-1} = - \hat{1}$ (of course, the Latin indices
$i,j,k ...$ should be raised and lowered with the metric $G^{ij}$ and its inverse). In the case of our
interest, the terms of orders higher than $\vp^2$ do not contribute
to the divergent part of the one-loop effective action and, therefore,
are not considered here.

It is worth noting that the latter feature is not true for other
models of quantum gravity. In fact, in the higher-derivative
fourth-order gravity only linear terms in $\varepsilon_i$ contribute
to the divergences~\cite{AvraBavi85,Avramidi-thesis}, while in
quantum general relativity in higher dimensions other terms are
necessary. For explicit expressions of the $O(\vp^3)$-terms, see
\cite{Cho:1991cja}. Calculations of the unique effective action in
$D\neq 4$ gravity models can be found, \textit{e.g.},
in~\cite{Huggins:1987zw,Cho:1991cja,Huggins:1986ht,Buchbinder:1988np_nj,Buchbinder:1989gy_qm,Buchbinder:1989iu}.
Even though we are mainly interested in
$D=4$ results, for the sake of generality, we let the space-time
dimension $D$ be arbitrary in our intermediate calculations.

\section{Field parametrizations and bilinear form of the action}
\label{Sec2}

In the traditional background field method the original field
$\,g_{\mu\nu}^\prime \,$  is split into a sum of a classical background
$\,g_{\mu\nu}\,$ and a quantum field $\,h_{\mu\nu}$, {\it i.e},
$\, g_{\mu\nu}^\prime \,=\, g_{\mu\nu} + \kappa h_{\mu\nu}$.
As in the present work we are interested in evaluating the one-loop
divergences in a general parametrization of the quantum field,
instead of performing the usual linear shift, we shall consider
$g_{\mu\nu}^\prime = f_{\mu\nu}(g_{\al\be},\phi_{\al\be})$.
Here, the indices are lowered and raised with the external metric
$g_{\mu\nu}$ (and its inverse $g^{\mu\nu}$) and $f$ depends on
the quantum field $\phi_{\mu\nu}$ possibly in a nonlinear way.
Assuming that $f$ has a series expansion, we can define the most
general (at one-loop order) parametrization of the quantum metric
in the form \cite{JDG-QG}
\beq
\n{t-exp}
g_{\mu\nu}^\prime
\, = \,
g_{\mu\nu} + \ka \, A_{(1) \, \mu\nu}^{\al\be} \, \phi_{\al\be}
+ \ka^2 A_{(2)\, \mu\nu}^{\la\tau, \rho \si}
\,\phi_{\la\tau} \phi_{\rho \si}
+ O(\ka^3),
\eeq
where $A^{...}_{(n)\,\mu\nu}$ are tensor structures depending only
on the background metric, and $\ka$ is the loop-expansion parameter.
Through covariance and symmetry arguments, the coefficient
functions in~\eq{t-exp} have the general tensor form
\begin{align}
\n{AA}
A^{\al\be}_{(1)\, \mu\nu} \,= & \,\, \ga_1 \, \de^{\al\be}_{\mu\nu}
+ \ga_2 \, g^{\al\be} g_{\mu\nu}\,,
\\
\begin{split}
\n{BB}
A^{\la\tau,\rho\om}_{(2)\, \mu\nu} \,= & \,\,
\frac{\ga_3}{2} \, g^{\ga \de} ( \de^{\la\tau}_{\ga(\mu} \de^{\rho\om}_{\nu)\de} + \de^{\rho\om}_{\ga(\mu} \de^{\la\tau}_{\nu)\de} )
+ \ga_4 \, \de^{\la\tau,\rho\om} g_{\mu\nu}
\\
&
\,\,
+ \frac{\ga_5}{2} \, ( \de^{\la\tau}_{\mu\nu} g^{\rho\om}
+ \de^{\rho\om}_{\mu\nu} g^{\la\tau} )
+ \ga_6 \, g^{\la\tau} g^{\rho\om} g_{\mu\nu}.
\end{split}
\end{align}
In these expressions
\beq
\de^{\mu\nu}_{\al\be} \,=\, \frac12 (\de^\mu_\al \de^\nu_\be
+ \de^\mu_\be \de^\nu_\al)
\label{1 metric}
\eeq
and $\ga_i$ ($i=1,\cdots ,6$) are six arbitrary coefficients
parametrising the choice of the quantum variable. The restrictions
$\,\ga_1 \neq 0\,$ and $\,\ga_1 + D \ga_2 \neq 0\,$
have to be imposed, to provide that the change of coordinates
from $g^\prime_{\mu\nu}$ to $\phi_{\mu\nu}$ is not degenerate. Terms
of order $O(\ka^3)$ in~\eq{t-exp} contribute only at the two- and
higher-loop orders and hence are irrelevant and will be omitted in what
follows. The one-loop contribution requires a functional integration
of a quadratic form in $\phi_{\mu\nu}$, hence it is evaluated
taking $\ka \to 0$ in Eq.~\eq{VEA}.

Inserting expressions \eq{AA} and \eq{BB} in Eq.~\eq{t-exp} we get
\begin{equation}
\begin{split}
\label{gen-bfm}
g_{\mu\nu}^\prime
=&\,\,
g_{\mu\nu}  + \ka \left( \ga_1 \phi_{\mu\nu}
+ \ga_2 \phi  g_{\mu\nu}\right)
\\
& + \,  \ka^2 \left(  \ga_3 \phi_{\mu\rho} \phi^\rho_\nu
+ \ga_4 g_{\mu\nu} \phi_{\rho\si} \phi^{\rho\si}
+ \gamma_5 \phi \phi_{\mu\nu}
+ \gamma_6 g_{\mu\nu} \phi^2 \right)  + O(\ka^3),
\end{split}
\end{equation}
where $\, g^{\mu\nu} \phi_{\mu\nu} \equiv \phi\,$ denotes the
trace of the quantum metric. The Eq.~\eq{gen-bfm} represents a
general parametrization of the quantum metric for one-loop
calculations.  Other choices of quantum variables based on the
expansions of
$\vert g^\prime \vert^p g^\prime_{\mu\nu}$ and
$\vert g^\prime \vert^q g^{\prime\mu\nu}$  (see, {\it e.g},
Refs.~\cite{Kalmykov:1995fd, Kalmykov:1998cv, Ohta:2016npm})
can be reduced to  particular cases of \eq{gen-bfm}. The explicit
values of  $\,\ga_i\,$ for these parametrizations are
displayed in Table~\ref{Table0}. Let us note that it is possible to
construct a parametrization of the more general type
$g_{\mu\nu}^\prime \,=\, e^{2\ka r \sigma} (g_{\mu\nu} + \cdots)$,
in which the conformal factor $\sigma(x)$ of the metric is explicitly
separated. Calculations using the conformal parametrization can be
found, {\it e.g.}, in \cite{JDG-QG,Kalmykov:1998cv,FirPei}. We
postpone the discussion on this choice to Sec.~\ref{ApB}.

\vskip 2mm
\begin{table}[h]
\centering
\begin{tabular}{|l|c|c|c|c|c|c|}
\hline
&
$\ga_1$ & $\ga_2$ & $\ga_3$ & $\ga_4$ & $\ga_5$ & $\ga_6$
\\
\hline
$ \vert g^\prime\vert^p g^\prime_{\mu\nu} $
& $\,\, \,\,1$ & $p$ & $0$ & $-p/2$ & $0$ & $ p^2/2 $
\\
\hline
$\vert g^\prime \vert^q g^{\prime\mu\nu}$
& $- 1 $ & $- q$ & $ 1 $ & $q/2$ & $ q$ &  $q^2/2$
\\
\hline
\end{tabular}
\begin{quotation}
\caption{\label{Table0} \sl Values of the parameters
in~\eq{gen-bfm} for the covariant and contravariant
densitized parametrizations.}
\end{quotation}
\end{table}
\vskip -4mm

The bilinear form of the action can be obtained by
expanding~\eqref{action} in powers of $\phi_{\mu\nu}$ by means
of~\eqref{gen-bfm}. This yields~\cite{JDG-QG}
\beq \label{action-phi}
S (g^\prime_{\mu\nu}) \,=\, S (g_{\mu\nu}) + S^{(1)} + S^{(2)}
+ \cdots\,,
\eeq
where
\begin{align}
\label{unili}
&
S^{(1)} \,=\, \frac{1}{\ka}  \int \text{d}^D x \sqrt{\vert g \vert}
\Big\{ \ga_1 R^{\mu\nu} \phi_{\mu\nu} - \tfrac12
\left[ \ga_1 + (D-2) \ga_2 \right]  R \phi
- (\ga_1 + D) \ga_2 \La \phi \Big\},
\\
\begin{split}
\label{bili}
&
S^{(2)} \,=\, - \frac12 \int \text{d}^D x \sqrt{\vert g \vert}
\, \Big\{
\phi_{\mu\nu} \left[  K^{\mu\nu,\al\be} (\Box - 2 \La)
+ M^{\mu\nu,\al\be}_1 + {M}_2^{\mu\nu,\al\be} \right]   \phi_{\al\be}
\\
&
\qquad
\qquad
+  ( \ga_1 \na_\rho \phi^{\rho}_\mu + \be \na_\mu \phi)^2 \,\Big\},
\end{split}
\end{align}
and unnecessary superficial terms have been omitted. In the last
formula
\beq
\be \,=\, - \frac12 \left[ \ga_1 + (D-2) \ga_2  \right]
\eeq
and the tensor objects are defined as
\begin{align}
K^{\mu\nu,\al\be}
=&\,\, \frac12 \Big\{ \ga_1^2 \de^{\mu\nu,\al\be}
- \frac{1}{2} \left[\ga_1^2 + 2(D-2) \ga_1 \ga_2
+ D(D-2) \ga_2^2 \right] g^{\mu\nu} g^{\al\be} \Big\},
\n{K-met}
\\
M^{\mu\nu,\al\be}_1 =&\,\,
 \ga_1^2 R^{\mu\al\nu\be} + \ga_1^2 g^{\nu\be} R^{\mu\al}
- \frac{x_1}{2} \, (g^{\mu\nu} R^{\al\be} + g^{\al\be} R^{\mu\nu})
- \frac{\ga_1^2}{2} \de^{\mu\nu,\al\be} R
+ \frac{x_2}{4} \, g^{\mu\nu} g^{\al\be} R,
\\
\begin{split}
{M}^{\mu\nu,\al\be}_2 =&\,\,
 - 2 \ga_3 g^{\nu\be} R^{\mu\al}
- \ga_5 (g^{\mu\nu} R^{\al\be} + g^{\al\be} R^{\mu\nu})
+ \left[\ga_3 + (D-2) \ga_4 \right] \de^{\mu\nu,\al\be} R
\\
&
\,\,
+ \left[\ga_5 + (D-2) \ga_6 \right] g^{\mu\nu} g^{\al\be} R
+ 2 (\ga_3 + D \ga_4) \de^{\mu\nu,\al\be} \La
+ 2 (\ga_5 + D \ga_6) g^{\mu\nu} g^{\al\be} \La,
\label{DeM}
\end{split}
\end{align}
with
\begin{equation}
\label{x1}
x_1 \,=\, \ga_1^2 + (D-4) \ga_1 \ga_2
\,,
\qquad \qquad
x_2 \,=\, \ga_1^2 + 2(D-4) \ga_1 \ga_2 + (D-2)(D-4) \ga_2^2.
\end{equation}
It is worth noticing that all the dependences on the parameters
$\ga_{3,\cdots,6}$ of the nonlinear part of the field
splitting~\eq{gen-bfm} are encoded in the tensor
${M}_2^{\mu\nu,\al\be}$.
In the above-given formulas, and in the following ones, we may
present  expressions in a compact form in which all algebraic
symmetries are implicit (for more details, see~\cite{JDG-QG}).

Finally, from Eq.~\eq{action-phi} it follows that the equations of
motion read
\beq
\n{EOM}
\vp^{\mu\nu} = \frac{1}{\sqrt{\vert g \vert}}
\frac{\de S}{ \de \phi_{\mu\nu}}
= \frac{1}{\ka} \, \big\{ \ga_1 R^{\mu\nu}
- \tfrac12 \big[\ga_1 + (D-2) \ga_2\big] R g^{\mu\nu}
- (\ga_1 + D) \ga_2 \La g^{\mu\nu} +  O(\ka)\big\}.
\eeq
Now, we have all basic elements to perform the desired calculation.

\section{Improved bilinear form of the action}
\label{Sec3}

General relativity and other metric theories of gravity are gauge
theories based on the diffeomorphism group $\mathscr{G}$. The
configuration space $\mathscr{M}$ is the set of all spacetime
metrics, and the coset $\mathscr{M}/\mathscr{G}$ is known as
the space of spacetime geometries. In quantum gravity
the invariant configuration-space metric is defined,
up to an arbitrary real parameter~$a$, by~\cite{DeWitt:1967yk}
\beq
\label{space-met-prime}
\de s^2
\,=\,
\int \text{d}^Dx \, \sqrt{\vert g^\prime \vert}
\, G^{\prime\mu\nu,\al\be} \de g_{\mu\nu}^\prime (x) \de g_{\al\be}^\prime(x),
\qquad
G^{\prime\mu\nu,\al\be}
\,=\,
\tfrac{1}{2} \, ( \de^{\prime\mu\nu,\al\be}
+ a g^{\prime\mu\nu} g^{\prime\al\be} ).
\eeq
The nondegeneracy of $\, G^{\prime\mu\nu,\al\be}\, $ is ensured
by the condition $ a \neq - 1/D$. Explicit calculations have shown
that the Vilkovisky effective action depends on the choice of
$a$~\cite{Huggins:1987zw,Odintsov:1991yx,BarberoG.:1993cw}.
The ambiguity owed to the parameter $a$ can be fixed by an
additional prescription.

A differential operator is said to be minimal if its highest-derivative
term is given only by a power of the $\Box$ operator. In
quantum gravity models, the minimal operator almost always has
the form of $G^{\mu\nu,\al\be}\Box^n$ with the parameter $a$
unambiguously
fixed by the choice of classical Lagrangian and the parametrization
of the quantum field. In Ref.~\cite{Vil-unicEA}, it was proposed
that $a$ should be chosen correspondingly; namely, the field-space
metric should be the expression in the highest-derivative term in
the minimal version of the bilinear part of the classical action.
This prescription relies on the assumption that
all the geometrical objects underlying the framework of
the unique effective action should be 
determined from the classical action~\cite{Vil-unicEA}.
For the quantum general relativity $n=1$ and, in the standard simplest
parametrization, this condition fixes the value $a \,=\, -1/2$. However,
even in the minimal gauge, the coefficient  of the term 
$g^{\mu\nu}g^{\al\be}$ of the field-space metric  may be changed by
modifying the parametrization of the quantum metric, that is, by changing
the coefficients $\ga_i$ in~\eqref{gen-bfm} (see, for instance, 
Eq.~\eqref{metric-0st} below).  One of the purposes of this
work is to check whether this change produces a modification
in the divergent part of the one-loop unique effective action. 

The field-space metric in terms of the variable $\phi_{\mu\nu}$
can be obtained by performing a change of variables in
Eq.~\eq{space-met-prime}, which gives
\beq
\de s^2 \, = \, \int \text{d}^D x \sqrt{\vert g \vert}
\, G^{\mu\nu,\al\be} \, \de \phi_{\mu\nu} (x) \de \phi_{\al\be} (x),
\eeq
where
\beq
\label{metric-all}
&&
G^{\mu\nu,\al\be} \,=\, G^{\mu\nu,\al\be(0)} + \ka \, G^{\mu\nu,\al\be(1)}
+ O(\ka^2),
\\
\label{metric-0st}
&&
G^{\mu\nu,\al\be(0)} \,=\, \frac12 (\ga_1^2 \de^{\mu\nu,\al\be}
+ \bar{a} \, g^{\mu\nu} g^{\al\be}),
\quad \quad
\bar{a} \,\equiv\,  \ga_2 (2 \ga_1 + D \ga_2) + a (\ga_1 + D \ga_2)^2,
\\
\label{metric-1st}
&&
G^{\mu\nu,\al\be(1)} \,=\,
  g_1 \, g^{\mu\al} \phi^{\nu\be}
+ g_2 \, {\de}^{\mu\nu,\al\be} \phi
+ g_3 \, (g^{\mu\nu} \phi^{\al\be} + g^{\al\be} \phi^{\mu\nu})
+ g_4 \, g^{\mu\nu} g^{\al\be} \phi,
\eeq
with the coefficients
\beq
&&
g_1  \,=\, - \ga_1^3 + 2 \ga_1 \ga_3,
\qquad
g_2  \,=\, \frac{\ga_1^2}{4} \left[\ga_1 + (D-4) \ga_2 \right]
+ \ga_1 \ga_5,
\nn
\\
&&
g_3  \,=\,
- \frac{\ga_1^2}{2}  \left[ 2 \ga_2
+ a \left(\ga_1 + D \ga_2 \right) \right]
+ \ga_2 \ga_3 + (\ga_1 + D \ga_2)[ \ga_4
+ a (\ga_3 + D \ga_4)] + \frac{\ga_1 \ga_5}{2},
\nn
\\
&&
g_4  \,=\, \frac{\bar{a}}{4} \, [\ga_1 +(D-4) \ga_2]
- \ga_1 \ga_2 [\ga_2 + a (\ga_1 + D \ga_2)]
\nn
\\
&&
\qquad
\qquad
+ \, 2 [\ga_1 \ga_6 + \ga_2 (\ga_5 + D \ga_6)
+ a (\ga_1 + D \ga_2) (\ga_5 + D \ga_6)].
\eeq

Formula~\eq{metric-0st} can be rewritten using the definition of
Eq.~\eqref{K-met},
\beq
\label{G0comK}
G^{\mu\nu,\al\be(0)}
\,=\, K^{\mu\nu,\al\be} +  \frac14 (1+2a)
(\ga_1 + D \ga_2)^2 g^{\mu\nu} g^{\al\be}.
\eeq
One can see that for $a = - 1/2$ the background configuration space metric
reduces to the factor of the d'Alembertian in~Eq.~\eqref{bili}.
This agrees with the Vilkovisky's prescription~\cite{Vil-unicEA}
for fixing the ambiguity in the one-parameter family of metrics,
even for the general parametrization~\eq{gen-bfm}.

The Christoffel symbol~\eqref{Gkij} associated with the
metric~\eqref{metric-all} has the form
\beq
\Ga^{ \mu\nu,\al\be }_{\rho\si} = \frac12 \, G_{\rho\si,\la\tau}
\Big(
\frac{\pa G^{\la\tau,\al\be} }{ \pa \phi_{\mu\nu} }
+ \frac{\pa G^{\mu\nu,\la\tau} }{ \pa \phi_{\al\be} }
- \frac{\pa G^{\mu\nu,\al\be} }{ \pa \phi_{\la\tau} }\Big)\,,
\label{Chris}
\eeq
where the inverse of the configuration-space metric \eq{metric-all} is
\beq
\label{42}
G_{\mu\nu,\al\be} \, = \,  K_{\mu\nu,\al\be}^{-1}
+  \frac{2(1+2a)}{(D-2)(1+aD)(\ga_1+D\ga_2)^2}
 \,\, g_{\mu\nu} g_{\al\be}
+ O(\ka)
\eeq
and $K_{\mu\nu,\al\be}^{-1}$ is the inverse of \eq{K-met},
\beq
&&
K_{\mu\nu,\al\be}^{-1} \,=\,
h_1 \de_{\mu\nu,\al\be}
+ h_2 \, g_{\mu\nu} g_{\al\be},
\n{K-1}
\\
\mbox{with}
&&
\qquad
\n{h12}
h_1 \,=\, \frac{2}{\ga_1^2},
\qquad
h_2 \, = \, - \frac{2}{D \ga_1^2} - \frac{4}{D(D-2)(\ga_1+D\ga_2)^2}.
\eeq
A straightforward calculation of \eqref{Chris} yields
\beq
\n{symbol}
\Ga^{\mu\nu,\al\be}_{\rho\si} \,=\, \ka  \big[
  c_1 \,  \de^{\mu\al}_{\rho\si} g^{\nu\be}
+ c_2 \, (\de^{\mu\nu}_{\rho\si} g^{\al\be}
  + \de^{\al\be}_{\rho\si} g^{\mu\nu})
+ c_3 \, {\de}^{\mu\nu,\al\be} g_{\rho\si}
+ c_4 \, g^{\mu\nu} g^{\al\be} g_{\rho\si}  \big]
+ O(\ka^2),
\eeq
where the coefficients are
\begin{align}
c_1 =& - \ga_1 + 2 \, \frac{\ga_3}{\ga_1},
\qquad
c_2 = \, \frac14 \, [ \ga_1 + (D-4) \ga_2 ] + \frac{\ga_5}{\ga_1},
\nn
\\
c_3 =& \, \frac{1}{2(D-2) (\ga_1 + D \ga_2)}
\left[ \ga_1^2 + 2(D-2) \ga_1 \ga_2
- \frac{(1+2a)D \ga_1^2}{2(1+aD)}
\right]
+ 2 \frac{\ga_1 \ga_4 - \ga_2 \ga_3 }{\ga_1(\ga_1 + D \ga_2)},
\nn
\\
c_4 =&  \,
- \frac{1}{4(D-2)(\ga_1 + D \ga_2) }
\left[ \ga_1^2 + 2(D-4) \ga_1 \ga_2 + (D-2)(D-4) \ga_2^2
- \frac{(1+2a)\ga_1^2}{(1+aD)}
\right]
\nonumber
\\
&\, +2  \frac{\ga_1 \ga_6 - \ga_2 \ga_5 }{\ga_1(\ga_1 + D \ga_2)}.
\nn
\end{align}

Using Eqs.~\eq{EOM} and \eq{symbol}, the Christoffel correction
term in the second covariant derivative
$\mathscr{D}_i \mathscr{D}_j S \,=\ S_{,ij} -  \Ga^k_{ij}\,\vp_k$
reads
\begin{equation}
\begin{split}
&
\Ga^{\mu\nu,\al\be}_{\rho\si} \varepsilon^{\rho\si} \big|_{\ka \to 0}
\,=\,
\frac{x_1}{4} (g^{\mu\nu} R^{\al\be} + g^{\al\be} R^{\mu\nu})
- \ga_1^2 \,   g^{\mu\al} R^{\nu\be}
+ \frac{\ga_1^2}{4} \, \de^{\mu\nu,\al\be} R
- \frac{x_2}{8}
\, g^{\mu\nu} g^{\al\be} R
\\
&
\,\, - {M}^{\mu\nu,\al\be}_2
+ \frac{D-4}{D-2} \, K^{\mu\nu,\al\be} \La
+ \frac{(1 + 2a)D\ga_1^2}{8(1+aD) } \,
  \Big( R + \frac{2D}{D-2} \, \La \Big)
  \big(\de^{\mu\nu,\al\be} - \frac{1}{D} \, g^{\al\be} g^{\mu\nu} \big),
\end{split}
\end{equation}
where $M_2^{\mu\nu,\al\be}$ and $x_{1,2}$ were defined in
Eqs.~\eq{DeM} and \eq{x1}, respectively. We remark that the
parameters $\ga_{3,..,6}$, which are related to the nonlinear terms
in the parametrization \eq{gen-bfm}, only occur in
$M^{\mu\nu,\al\be}_2$, just as in~\eqref{bili}. Because of
this, the second functional covariant derivative of the
action~\eqref{action-phi} only depends on the parameters $\ga_1$
and $\ga_2$,
\beq
&&
-\frac{\mathscr{D}^2 S}{ \de \phi_{\mu\nu} \de \phi_{\al\be}}
\Bigg|_{\ka \to 0}
\,=\,
\frac{\ga_1^2}{2} \de^{\mu\nu,\al\be} \Box
- \frac{d_1}{2}  g^{\mu\nu} g^{\al\be} \Box
+ \frac{d_2}{2} (g^{\mu\nu} \na^{\al} \na^\be + g^{\al\be} \na^{\mu} \na^\nu)
\nn
\\
&&
\qquad
- \, \ga_1^2 g^{\mu\al} \na^{\nu\be}
+ \ga_1^2 R^{\mu\al\nu\be} 
- \frac{x_1}{4} (g^{\mu\nu} R^{\al\be} + g^{\al\be} R^{\mu\nu} )
- \frac{\ga_1^2}{4} \, \de^{\mu\nu,\al\be} R
+ \frac{x_2}{8} g^{\mu\nu} g^{\al\be} R
\nn
\\
&&
\qquad
-\, \frac{D}{D-2} \, K^{\mu\nu,\al\be} \La
+ \frac{(1 + 2a)D\ga_1^2}{8(1+aD) } \,
\Big(\de^{\mu\nu,\al\be} - \frac{1}{D}\, g^{\al\be} g^{\mu\nu} \Big)
\Big( R + \frac{2D}{D-2} \, \La \Big),
\label{2covS}
\eeq
where
\begin{equation}
d_1 \,=\, \ga_1^2 + 2(D-2) \ga_1 \ga_2 + (D-1)(D-2) \ga_2^2
, \qquad \,\,
d_2 \,=\, \ga_1^2 + (D-2) \ga_1 \ga_2.
\end{equation}

It is clear that the Christoffel symbol derived from the
metric~\eqref{metric-all} should suffice to compensate the
dependence of $S_{,ij}$ on the nonlinearity of the field
parametrization. In fact, for $\ka \to 0$  all the parameters
$\,\ga_{3,\, \cdots,\,6}\,$ only contribute to the last term in
the {\it r.h.s.}~of
\beq
\frac{\de^2 S'}{\de g^\prime_{\mu\nu} \de g^\prime_{\al\be}} \,=\,
\frac{\de \phi_{\la\tau}}{\de g^\prime_{\mu\nu}}
\frac{\de \phi_{\rho\si}}{\de g^\prime_{\al\be}}
\frac{\de^2 S}{\de \phi_{\la\tau} \de \phi_{\rho\si}}
+
\frac{ \de^2 \phi_{\la\tau} } {\de g^\prime_{\mu\nu} \de g^\prime_{\al\be}}
\frac{\de S}{\de \phi_{\la\tau}}
\,,
\eeq
which represents the non-tensor nature of this transformation.

\section{One-loop divergences of Vilkovisky effective action}
\label{Sec4}

Up to this point, we have considered the part of the
Vilkovisky effective action based on the Christoffel symbols on the
space $\mathscr{M}$ of field parametrization. However, it is still
necessary to introduce the gauge fixing for the diffeomorphism
invariance and take into account the contribution of the Faddeev-Popov
ghosts as well the terms~\eqref{U1op} and~\eqref{U2t}
related to the gauge constraints on the affine connection.

The standard general form of the gauge-fixing action in quantum
general relativity is
\beq
\label{S_gf}
S_{\text{GF}} \,=\,
\frac12 \int \text{d}^Dx \sqrt{\vert g \vert}
\,\chi_\al g^{\al\be} \chi_\be,
\eeq
where $\chi_\al$ is the background gauge condition.
The use of a linear gauge fixing\footnote{See Ref.~\cite{GLS19}
for a recent discussion on nonlinear gauges within the
framework of the background field method in the standard
definition of the effective action.} is not a necessary condition to
ensure the invariance of the Vilkovisky effective
action~\cite{Fradkin:1983nw,Rebhan:1986wp}. Nonetheless, as
explained in Sec.~\ref{Sec1}, the DeWitt
gauge~\eqref{de-gauge} is crucial for deriving the expanded
formula~\eq{VEA}. In our parametrization, it assumes the form
\beq \label{DeWittGauge}
\chi_\al \, = \,  G^{\mu\nu,\la\tau} \, R_{\mu\nu,\al} \, \phi_{\la\tau}
\, = \,  - \ga_1 \na_\rho \phi^{\rho}_\al -
\left[\ga_2 + a \left(\ga_1 + D \ga_2 \right) \right] \na_\al \phi + O(\ka),
\label{DeWittGaugeB}
\eeq
where we used the explicit expression for the generators of the gauge
transformations $R_{\mu\nu,\al}$ of the field $\phi_{\mu\nu}$,
presented in the Appendix.

Comparing Eqs.~\eqref{DeWittGaugeB} and~\eq{bili} it is easy to see
that the choice $\, a = -1/2 \,$ provides the minimal form of the operator~\eq{H},
\beq
\label{Hexp}
\hat{H}
\,=\,   G_{\mu\nu,\rh\si}
\Big(
\frac{\mathscr{D}^2 S}{ \de \phi_{\rh\si} \de \phi_{\al\be}}
+ \frac{\de \chi_{\la}}{\de \phi_{\rh\si}}
\, g^{\la\ta} \frac{\de \chi_{\ta}}{\de \phi_{\al\be}} \Big)
\Big|_{\ka \rightarrow  0}.
\eeq
Let us remark that another possible way of making the operator
$H^{\mu\nu,\al\be}$ minimal is through the use of a specific parametrization,
namely, $\ga_1 = - D \ga_2$. However, as explained in Sec.~\ref{Sec2},
this is not acceptable since it makes the metric in the space of the
quantum fields singular, see Eq.~\eqref{42}, and the operator $\hat{H}$
in~\eqref{Hexp} undefined. Thus, $a = -1/2\,$ is the sole reasonable choice.
For this value of $a$, the operator gets reduced to the standard form
\beq
\label{H_final}
\hat{H}\,=\, -\, (\hat{1} \Box + \hat{\Pi}),
\eeq
where $\hat{1} \,=\, \de^{\mu\nu}_{\al\be}$ is the identity
operator~\eqref{1 metric} on the space of symmetric rank-2 tensors and
\beq
  \hat{\Pi} = 2  R{}^{\mu}_{\,.\,\al}{}^{\nu}_{\,.\,\be}
- \frac{p_1}{2} \, g^{\mu\nu} R_{\al\be}
- \frac{p_2}{D-2} \, g_{\al\be} R^{\mu\nu}
+ \frac{p_3}{2(D-2)} \, g^{\mu\nu} g_{\al\be} R
+ \de^{\mu\nu}_{\al\be} \Big(\frac{D\La}{D-2} - \frac12 \,R\Big),
\eeq
with
\beq
&&
p_1 = 1 + \frac{\ga_2(D-4) }{\ga_1}
,\quad
p_2 = \frac{\ga_1 + 2 (D-2) \ga_2} {\ga_1 + D \ga_2}
,\quad
p_3 = p_2 +
\frac{(D-2) (D-4) \ga_2^2}{\ga_1 (\ga_1 + D \ga_2) }.
\nn
\eeq

Furthermore, with the gauge condition~\eqref{DeWittGaugeB},
the ghost matrix reads
\beq
\label{N_ab}
\hat{N} =  g^{\al\la} G^{\mu\nu,\rh\si} R_{\mu\nu,\la}  R_{\rh\si,\be}
=  \de^{\al}_\be \Box + (1 + 2a) \na^\al \na_\be + R^\al_\be   + O(\ka)
\,.
\eeq
Notice that in the DeWitt gauge all the dependence on the
parametrization is cancelled in the ghost operator, and
that $a = -1/2$ makes it also minimal.
Hereafter, we choose this value for $a$, such that both $\hat{H}$
and $\hat{N}$ assume minimal forms.

The correction which is responsible to restore the gauge invariance
of the effective action is based on the nonlocal operators
$\,\hat{U}_1$ and $\, \hat{U}_2$, defined in \eq{U1op} and \eq{U2t}.
These operators depend on the two new vertices
\beq
\label{V1eV2}
(V_1)_{i\al} \,=\, (\mathscr{D}_i  R^j_\al) \, \vp_j
\qquad \text{and} \qquad
(V_2)_{\al\be} = R^{i}_\al \, (\mathscr{D}_i R^j_{\be}) \, \vp_j.
\eeq
Particularizing the formulas above for the gravity theory in the
parametrization~\eqref{gen-bfm} and using the gauge
generators~\eq{ge-phi} given in Appendix, after some
algebra we get
\begin{equation}
\label{V1}
\begin{split}
(V_1)^{\mu\nu}_{\ga} \,=&\,\,
\frac{\ga_1}{2} \, ( R^\mu_\ga \na^\nu + R^\nu_\ga \na^\mu )
-  \frac{\ga_1}{2}  \, (\de^{\mu}_{\ga} R^{\nu\la}
+ \de^\nu_\ga R^{\mu\la}) \na_\la
+ \ga_1 \, (\na_\ga R^{\mu\nu})
\\
&
+ \frac{\ga_1}{2} \, R^{\mu\nu} \na_\ga
- \frac12 (\ga_1 + D \ga_2) \, g^{\mu\nu} R^\la_\ga \na_\la
+ \frac{\ga_1}{4}  R ( \de^\mu_\ga \na^\nu + \de^\nu_\ga \na^\mu )
\\
&
- \frac12 [\ga_1 + (D-2) \ga_2] \, g^{\mu\nu} (\na_\ga R)
- \frac14 [\ga_1 + (D-4) \ga_2] \, g^{\mu\nu} R \, \na_\ga
\\
&
+ \frac{D \ga_1}{ 2(D-2)} \La ( \de^\mu_\ga \na^\nu + \de^\nu_\ga \na^\mu )
- \frac{D [\ga_1 + (D-2) \ga_2]}{ 2(D-2)} \, g^{\mu\nu} \La \, \na_\ga
+ O(\ka)
\end{split}
\end{equation}
and
\begin{equation} \label{V2}
\begin{split}
(V_2)_{\al\be} =& \,\,
    R_{\al\be} \Box
    + \frac12 \, g_{\al\be} R \Box
-  g_{\al\be} R^{\la\tau} \na_\la \na_\tau
+  (\na^\la R_{\al\be} ) \na_\la
 - (\na_\al R^\la_\be) \na_\la
+(\na_\be R^\la_\al) \na_\la
\\
&
-   R_{\al\la\be\tau} R^{\la\tau}
+   R_{\al\la} R^{\la}_{\be}
+ \frac12  R R_{\al\be}
+ \frac{D\La}{D-2} \, (g_{\al\be} \Box + R_{\al\be})
+ O(\ka)
\,.
\end{split}
\end{equation}
We see that the dependence on the parameters $\ga_{3,...,6}$
corresponding to the nonlinear part of the field splitting~\eqref{gen-bfm}
gets cancelled in $(V_1)^{\mu\nu}_{\ga}$, while the vertex
$(V_2)_{\al\be}$ is parametrization independent automatically.

The operators  $\hat{U}_1$ and  $\hat{U}_2$ can be obtained by
substituting the two previous equations into the formulas~\eq{U1op}
and~\eq{U2t}, together with the propagators
\beq
\label{Ninv}
N^{\al\be} \,=\,  - g^{\al\be}\frac{1}{\Box}
+ R^{\al\be} \frac{1}{\Box^2} + O([m]^3)\,,
\qquad
H^{-1}_{\mu\nu,\al\be}
\,=\, K^{-1}_{\mu\nu,\al\be} \, \frac{1}{\Box} + O([m]^2).
\eeq
Here, $O([m]^k)$ denotes a series of inessential terms of higher
background dimension $k$. Remember that, according to
\cite{Bavi83-85}, for a functional universal trace
\beq
\Tr \hat{C}^{\mu_1 \cdots \mu_k} \na_{\mu_1}
\cdots \na_{\mu_k} \,\frac{\hat{1}}{\,\,\Box^n}\,,
\eeq
the background dimension (in mass units) is defined as the
dimension of the tensorial coefficient
$\hat{C}^{\mu_1 \cdots \mu_k}$, and its
superficial degree of divergence is expressed by the relation
$\om \,=\, D  - 2n + k$. Thus, in four dimensions only the traces
with background dimension 0, 1, 2, 3 and 4 contribute to the
ultraviolet (UV) divergences.

With all these ingredients in hand, it is possible to evaluate the
contribution of each term in~\eq{VEA}, up to background dimension
$O([m]^4)$, to the effective action. In the case of the operators
$\hat{H}$ and $\hat{N}$ (respectively given by Eqs.~\eqref{H_final}
and~\eqref{N_ab}), this can be obtained from the functional trace of
the coefficient $\hat{a}_2$ of the
Schwinger-DeWitt expansion~\cite{BDW-65}. On the other hand,
the functional traces of the nonlocal operators $\hat{U}_1$,
$\hat{U}_1^2$ and $\hat{U}_2$ can be evaluated using the table
of universal functional traces within the generalized
Schwinger-DeWitt technique~\cite{Bavi83-85}. For example, one
can easily show that
\beq
\Tr \hat{U}_2 \,=\,
\int \text{d}^D x \, \tr \left[  h_1 (V_1^2)^{\al}_\be  +
h_2 (\bar{V}_1^2)^{\al}_\be \right]
\frac{1}{\,\Box^3} \Big|_{x^\prime \to x } \,+\, O([m]^5),
\eeq
where $\,h_{1,2}\,$ were defined in Eq.~\eq{h12} and we used the
notations
\beq
(V_1^2)^{\al}_\be = g^{\al\ga} \de_{\mu\nu,\rh\si} (V_1)^{\mu\nu}_{\ga}
(V_1)^{\rh\si}_{\be},
\quad
(\bar{V}_1)_\ga = g_{\mu\nu} (V_1)^{\mu\nu}_{\ga},
\quad
(\bar{V}_1^2)^{\al}_\be = g^{\al\ga} (\bar{V}_1)_\ga (\bar{V}_1)_\be.
\nn
\eeq

Skipping the algebra, the contributions of the
terms in \eq{VEA} to the $\frac{1}{D-4}$-pole of the Vilkovisky
unique effective action is presented in  Table~\ref{Table2}. It is
important to recall that only in $D \to 4$ the displayed coefficients
correspond to one-loop divergences; nonetheless, our calculation
in arbitrary dimension shows that they do not depend on the field
parametrization even for $D\neq 4$. Moreover, one can see that the
parametrization dependence which remained after the Christoffel
correction was taken into account is cancelled in the functional trace
of each operator on its turn, as none of the coefficients depends on
$\ga_{1,2}$.


\begin{table}[ht]
\centering
\begin{footnotesize}
\begin{tabular}{|l|ccccc|c|}
\hline
Invariant
&
$\frac{i}{2} \Tr \ln \hat{H}$ &$ -i\Tr \ln \hat{N} $ &$- \frac{i}{2}\Tr \hat{U}_1$ &$- \frac{i}{4}\Tr \hat{U}_1^2$ &$ \frac{i}{2}\Tr \hat{U}_2$ & $\bar{\Ga}^{(1)}$
\\
\hline
$ R_{\mu\nu\al\be}^2$
& $\frac{D^2 - 29D + 480}{360} $ & $\frac{15-D}{90}$ & $0$ & $0$ & $0$ & $ \frac{D^2-33D+540}{360} $
\\
$R_{\mu\nu}^2 $
& $- \frac{D(D^2 - D + 178)}{360(D-2)} $ & $\frac{D-90}{90}$ & $ \frac{D + 12}{6} $ & $\frac{D+12}{24}$ & $ -\frac{3D^2-16}{8(D-2)}$ &
$- \frac{D^3 +55 D^2 -204 D + 360}{360 (D-2)} $
\\
$R^2$
& $\frac{D^3 - D^2 + 10 D - 6}{36(D-2)} $ & $-\frac{D+12}{36}$ & $\frac{1}{6}$ & $ \frac{D+12}{48}$ &  $ -\frac{3D-4}{8(D-2)}$ &
$\frac{4 D^3 - 5D^2 + 24}{144 (D-2)}$
\\
$\La R$
& $\frac{D(D^2+D+6)}{6(D-2)}$ & $0$ & $\frac{D(D+6)}{6(D-2)}$ & $ \frac{D(D+4)}{4(D-2)}$ & $-\frac{D (D + 4)}{2 (D - 2)}$ & $\frac{D(2D^2+D+12)}{12(D-2)}$
\\
$\La^2$
& $\frac{D^3 (D+1)}{4(D-2)^2}$ & $0$ & $0$ & $ \frac{D^3}{2(D-2)^2}$ & $-\frac{D^3}{(D-2)^2}$ & $\frac{D^3(D-1)}{4(D-2)^2}$
\\
\hline
\end{tabular}
\begin{quotation}
\caption{\label{Table2} \sl Contribution of each operator
in~\eqref{VEA} to the coefficients of each curvature invariant
in the divergent  (at $D\to 4$) part of the one-loop Vilkovisky
effective action. Each invariant enters the effective action multiplied
by the overall coefficient as in Eq.~\eqref{Final}. The final
coefficients, which are the sum of the coefficients of columns
2--6, are presented in the last column.}
\end{quotation}
\end{footnotesize}
\end{table}

Since the object of our interest is the one-loop logarithmically
divergent part of the Vilkovisky effective action, in the framework
of dimensional regularization we can take the limit $D \to 4$ in
the coefficient of the pole term, to obtain
\beq
\bar{\Ga}^{(1)}_{\text{div}}
\,=\,  - \frac{\mu^{D-4}}{(4\pi)^2 (D-4)} \int \text{d}^D x \sqrt{\vert g \vert}
\left\{ \frac{53}{45} R_{\mu\nu\al\be}^2
-\frac{61}{90} R_{\mu\nu}^2 + \frac{25}{36} R^2 + 8 \La R + 12 \La^2
\right\}.
\label{Final}
\eeq
As usual, $\mu$ is the renormalization parameter.
Formula \eq{Final} reproduces the results for the Vilkovisky effective
action for general relativity calculated in the standard, particular,
parametrization of the quantum variables in~\cite{Bavi83-85} (the
coefficients of the terms related to the cosmological constant were
calculated for the first time in~\cite{Fradkin:1983nw}). Moreover,
it is straightforward to verify that, on the classical mass shell, the
divergences of Eq.~\eqref{Final} correctly reduce to the coefficients
of the usual on-shell effective action~\cite{hove,chrisduff},
\beq
\bar{\Ga}^{(1)}_\text{div} \big|_{\text{on-shell}}
\,=\, - \frac{\mu^{D-4}}{(4\pi)^2 (D-4)} \int \text{d}^D x \sqrt{\vert g \vert}
\left\{
\frac{53}{45} R_{\mu\nu\al\be}^2 -\frac{58}{5}  \La^2
\right\}.
\eeq
This is an expected result since the Vilkovisky correction term is
proportional to the equations of motion. On the other hand, this result
is known to be gauge-fixing and parametrization independent
\cite{JDG-QG}.

It is interesting to compare the result for the unique effective
action~\eqref{Final} and the one-loop divergences of the standard (usual)
effective action in an arbitrary parametrization~\eqref{gen-bfm},
derived in \cite{JDG-QG}. It turns out that the two
expressions coincide if the parameters satisfy the conditions
\beq
\ga_4 & = & \frac{1}{48} \left[ \left( 6 \pm \sqrt{15} \right) \ga_1^2
- 12 \ga_3 \right] ,
\\
\ga_5 & = & \frac{1}{12} \left[ - 6 \ga_3 \pm
\left( 1 + \frac{4\ga_2}{\ga_1} \right)
\sqrt{6 \left( 12\ga_3^2 - 5 \ga_1^4 \right) } \right] ,
\label{rel2}
\\
\ga_6 & = & -\frac{1}{64}
\left[ 5 \left( \ga_1 + 4 \ga_2 \right)^2
+ 4 \left[  \ga_3 + 4 \left( \ga_4 +  \ga_5 \right) \right]  \right].
\eeq
In this case, the one-loop divergences of the conventional effective
action calculated in the minimal gauge coincide with those of the
Vilkovisky effective action~\eqref{Final}. Curiously, this result
can be achieved only if the parametrization is nonlinear. This 
can be readily seen from Eq.~\eqref{rel2}, which implies
$\ga_3 \neq 0$. Let us note that the observation formulated above can
be seen as a parametrization-dependence counterpart for the result
of~\cite{Lavrov:1988is}, where it was derived a \textit{gauge} for
which the one-loop divergences of the conventional effective
action (in the particular simplest parametrization) reproduce
those of the unique effective action. In this vein, it is also worth
pointing out that the $\Lambda$-dependent terms in~\eqref{Final}
can be obtained by means of the Landau-DeWitt gauge within the usual
definition of the effective action~\cite{Fradkin:1983nw}. Nevertheless,
the simple use of this particular singular gauge in the standard
effective action cannot give the other divergent terms of the unique
effective action for Einstein gravity because the space of fields is
not flat~\cite{Fradkin:1983nw,Rebhan:1986wp,Ellicott:1987ir}.

\section{Conformal parametrization of the metric}
\label{ApB}

Let us now consider a more general parametrization of the metric,
which explicitly splits its conformal factor, namely,
\begin{equation}
\begin{split}
\label{AppBEq1}
g_{\mu\nu}^\prime
=& \,\,
e^{2 \ka r \sigma} \big[ g_{\mu\nu}
+ \kappa (\gamma_1 \phi_{\mu\nu} + \gamma_2 \phi g_{\mu\nu} )
\\
&+\,
\kappa^2 (\gamma_3 \phi_{\mu\rho} \phi^\rho_\nu
+ \gamma_4 g_{\mu\nu} \phi_{\rho\sigma}^2
+ \gamma_5 \phi \phi_{\mu\nu} + \gamma_6 \phi^2 g_{\mu\nu} )
+ O(\ka^3) \big],
\end{split}
\end{equation}
where $\,g_{\mu\nu}\,$ is the background metric, $\phi_{\mu\nu}$
and $\si$ are the quantum fields and $\ga_{1,\cdots,6}$ and $r$ are
arbitrary parameters. The one-loop divergences of the standard
effective action for Einstein gravity were evaluated in this
parametrization in Ref.~\cite{JDG-QG}.

It turns out that it is not possible to construct the Vilkovisky
effective action directly in this parametrization. Treating the
conformal factor $\si$ as a new field increases the total number of
scalar modes. As a consequence, there is an artificial conformal
symmetry and related degeneracy, making the transformation singular.
For instance, the metric in the configuration space is
\beq
{ G}^{AB}
\label{defmet}
= \begin{pmatrix}
G^{\mu\nu,\al\be(0)}  & r(\ga_1+D\ga_2)(1+aD) g^{\mu\nu} \\
r(\ga_1+D\ga_2)(1+aD) g^{\al\be} & 2 r^2 D (1 + a D )
\end{pmatrix} + {O} (\ka) ,
\label{MetDegen}
\eeq
where $A,B,\cdots$ take the labels $\phi_{\mu\nu}$, $\si$, and
$G^{\mu\nu,\al\be(0)}$ coincides with Eq.~\eqref{G0comK}.
The determinant of the $O(\ka^0)$-term of this metric reads
\beq
\big| {G}^{AB(0)} \big|
\,=\,
\left\{ 2 r^2 D (1 + a D )
\,-\,  r^2(\ga_1+D\ga_2)^2(1+aD)^2
g^{\mu\nu} g^{\al\be} G_{\mu\nu,\al\be}^{(0)} \right\}
\times
\big| G^{\mu\nu,\al\be(0)}\big|.
\label{det}
\eeq
It is straightforward to verify that the term in curly brackets vanishes,
proving that the field-space metric is degenerate.
Therefore, it is not possible to evaluate the Christoffel symbols.
The problem originates from the fact that the change of variables
$g_{\mu\nu}^\prime \mapsto ( \phi_{\mu\nu}, \si )$ is not a
diffeomorphism.

The described difficulty can be resolved as
follows. We impose, from the
beginning, the additional conformal gauge fixing
\beq
\n{cGF}
\si = \la \phi
\eeq
with $\la$ being the gauge-fixing parameter. Expanding the
exponential in~\eqref{AppBEq1} one can see that, up to order
$\kappa^2$, this parametrization reduces to~\eqref{gen-bfm} via
the substitutions
\beq
\ga_2 \longmapsto \ga_2  + 2r\lambda,
\qquad
\ga_5 \longmapsto \ga_5 + 2r \lambda \ga_1,
\qquad
\ga_6 \longmapsto \ga_6 + 2r \lambda \ga_1.
\eeq
Then, all calculations that we carried out for~\eqref{gen-bfm}
also apply for the conformal parametrization~\eqref{AppBEq1}.

An alternative approach is to split the field $\phi_{\mu\nu}$ in
the trace and traceless part, that is,
\beq \label{84}
\n{tt-t}
\phi_{\mu\nu} = \bar{\phi}_{\mu\nu} + \frac{1}{D}\, g_{\mu\nu}\phi .
\eeq
It is clear that $g^{\mu\nu} \bar{\phi}_{\mu\nu} = 0$. We now have a
parametrization in terms of two independent quantum fields:
$\bar{\phi}_{\mu\nu}$ and $\phi$.
Applying~\eqref{cGF} and~\eqref{84} in~\eqref{AppBEq1} we get
\beq
&&
g_{\al\be}^\prime  =  g_{\alpha\beta}
+ \kappa ({\gamma}_1 \bar{\phi}_{\alpha\beta}
+ \bar{\gamma}_2 \phi g_{\alpha\beta} )
\nn
\\
&&
\quad
+\, \kappa^2 ({\ga}_3 \bar{\phi}_{\al\rho} \bar{\phi}^\rho_\be
+ {\ga}_4 \bar{\phi}_{\rho\si} \bar{\phi}^{\rho\si} g_{\al\be}
+ \bar{\gamma}_5 \phi \bar{\phi}_{\alpha\beta}
+ \bar{\gamma}_6 \phi^2 g_{\alpha\beta} ) + O(\ka^3),
\eeq
where the new coefficients are
\begin{align}
\bar{\gamma}_2 &= \frac{\ga_1}{D} + \ga_2 + 2 r \la,
\nn
\\
\bar{\gamma}_5 &= \frac{2\ga_3}{D} + \ga_5 + 2 \ga_1 r \la,
\nn
\\
\bar{\gamma}_6 &= \frac{1}{D^2} \left[ \ga_3
+ D (\ga_4 + \ga_5) + D^2 \ga_6
+ 2 D \left( \ga_1 + D \ga_2 \right) r \la  \right]  + 2 r^2 \la^2.
\end{align}
Now it is possible to define a nonsingular metric in the space of
the fields\footnote{Here, to avoid any kind of ambiguity, we made
use of a more explicit notation for the indices.},
\beq
&&
{G}^{ \bar{\phi}_{\mu\nu},\, \bar{\phi}_{\al\be} }
= \ga_1^2 \bar{\delta}^{\mu\nu,\al\be}
+ \ka \big[\zeta_1 g^{\mu\al} \bar{\phi}^{\be\nu}
+ \zeta_2 \bar{\delta}^{\mu\nu,\al\be} \phi \big]
+ O(\ka^2),
\nn
\\
&&
{G}^{\bar{\phi}_{\al\be} , \, \phi }
=  \ka \, \zeta_3 \, \bar{\phi}^{\al\be} + O(\ka^2) ,
\nn
\\
&&
{G}^{\phi,\, \phi}
=   \bar{\ga}_2^2 D (1+aD) + \ka \,  \zeta_4 \,  \phi + O(\ka^2) ,
\eeq
where
$\,\bar{\delta}^{\mu\nu}_{\al\be} = \delta^{\mu\nu}_{\al\be}
- \frac{1}{D} g^{\mu\nu}g_{\al\be}\,$
is the identity operator in the space of traceless symmetric rank-2
tensors, and the coefficients read
\begin{align}
\zeta_1 & \,=\,   - 2 \ga_1 ( \ga_1^2 - 2  \ga_3 ),
\nn
\\
\zeta_2 &\,=\,\frac{D-4}{2}\,\ga_1^2 \bar{\ga}_2+2 \ga_1\bar{\ga}_5,
\nn
\\
\zeta_3 & \,=\,  2 \bar{\ga}_2 ( 1 + a D ) ({\ga}_3
+ D {\ga}_4 ) + \ga_1 \bar{\ga}_5 - \ga_1^2 \bar{\ga}_2 (2 + a D ),
\nn
\\
\zeta_4  & \,=\,  \bar{\ga}_2 D ( 1 + a D ) \Big( \frac{D-4}{2}
\, \bar{\ga}_2^2 + 4 \bar{\ga}_6  \Big).
\end{align}
The inverse metric
$( { G}^{-1})_{AB}$ ($A,B, \cdots = \bar{\phi}_{\mu\nu}, \phi$)
is given by
\beq \label{InvMat}
({ G}^{-1})_{AB} =
\begin{pmatrix}
\frac{1}{\ga_1^2} \bar{\delta}_{\mu\nu,\al\be} & 0 \\
0 & \frac{1}{\bar{\ga}_2^2 D (1 + a D )}
\end{pmatrix} + {O} (\ka) .
\eeq

With these ingredients, we can proceed with the evaluation of the
Christoffel symbols, whose non-zero components are
\beq
&&
{\Ga}^{\bar{\phi}_{\mu\nu},\,\bar{\phi}_{\al\be}}_{\bar{\phi}_{\la\ta}}
=
\frac{\ka \ze_1}{ \ga_1^2} \,  g^{\mu\al} \bar{\delta}^{\be\nu}_{\la\ta}
+ O(\ka^2),
\nn
\\
&&
{\Ga}^{ \bar{\phi}_{\mu\nu}, \, \bar{\phi}_{\al\be}} _{\phi}
= \ka \Big[\,
\frac{2 (\ga_3 + D \ga_4)}{D \bar{\ga}_2}
-  \frac{\ga_1^2 (4 + D + 4 a D )}{4 D ( 1 + a D ) \ga_2}\Big]
\bar{\de}^{\mu\nu,\al\be} + O(\ka^2),
\nn
\\
&&
{\Ga}^{\bar{\phi}_{\mu\nu},\, \phi}_{\bar{\phi}_{\la\ta}}
= \ka \Big( \frac{D-4}{4}\ \bar{\ga}_2
+ \frac{\bar{\ga}_5}{\ga_1} \Big)  \bar{\de}^{\mu\nu}_{\la\ta} + O(\ka^2),
\nn
\\
&&
{\Ga}^{\phi,\phi}_{\phi}
= \ka \Big( \frac{D-4}{4} \, \bar{\ga}_2
+ \frac{2 \bar{\ga}_6}{\bar{\ga}_2} \Big) + O(\ka^2).
\eeq
For the second covariant derivative of the action we have
\beq
&&
\frac{\mathscr{D}^2 S}{\de \bar{\phi}_{\mu\nu}
\de \bar{\phi}_{\al\be}}\bigg|_{\ka \to 0}
= \ga_1^2 \Big[ g^{\be\nu} \nabla^\al \nabla^\mu
- \frac{1}{2} \bar{\de}^{\mu\nu,\al\be} \Box - R^{\mu\al\nu\be}
- \frac{1}{4(1+aD)} \Big(
\frac{D-2}{2} R  + D \La   \Big) \bar{\de}^{\mu\nu,\al\be}
\Big],
\nn
\\
&&
\frac{\mathscr{D}^2 S}{\de \bar{\phi}_{\mu\nu} \de \phi}
\bigg|_{\ka \to 0}
=  \ga_1 \bar{\ga}_2
\Big(   -\frac{D-2}{2} \, \nabla^\mu \nabla^\nu
+  \frac{D-4}{4} R^{\mu\nu} \Big),
\nn
\\
&&
\frac{\mathscr{D}^2 S}{\de\phi \de\phi}\bigg|_{\ka \to 0}
=   \bar{\ga}_2^2 \Big[ \frac{(D-2)(D-1)}{2}  \Box
- \frac{(D-4)(D-2)}{8}R
- \frac{D^2}{4} \La \Big].
\label{OffDiag}
\eeq

At this stage, it is clear that the dependence on the nonlinear quantum
field parametrization was compensated by the Christoffel correction,
just like in~\eqref{2covS}. In addition, the use of the parametrization
in terms of the traceless and trace parts reveals that the improved
bilinear operator can be written as constant matrix times a differential
operator independent of $\ga_1$ and $\bar{\ga}_2$; thus, this dependence
is trivial.

We point out that the conformal gauge fixing \eq{cGF} does not
require Faddeev-Popov ghosts because the conformal transformation
has no derivatives \cite{frts82}. Moreover, under the
diffeomorphism~\eqref{CoorTrans} the field $\si$ transforms as
$\delta \si = - \na_\mu \si \, \xi^\mu$, and the terms in the ghost operator
associated with the generators $R_\mu = - \na_\mu \si$ can be safely
ignored at one-loop level since they produce third-order contributions in
quantum field; as a consequence, we get~\eqref{N_ab}. Therefore,
even in the conformal parametrization,
the final result matches the one presented in Eq.~\eq{Final} once
the conformal factor is identified with the trace of $\phi_{\mu\nu}$.

\section{Conclusions}
\label{Sec5}

We performed the calculations of the one-loop divergences of the
Vilkovisky unique effective action in quantum general relativity
in an arbitrary, most general
parametrization of quantum metric, including the conformal
parametrization and the corresponding gauge fixing. Because of the
similarity between the conformal parametrization and the two-dimensional
quantum gravity, one could suspect that the unique effective action
may lose its invariance and universality.
We have shown that this
does not happen and the one-loop divergences are universal.
In order to achieve the positive result in the excessive conformal
parametrization, the conformal gauge should be fixed before applying
Vilkovisky's formalism, to guarantee the non-degeneracy of the
field-space metric.

Finally, we fixed the dependence of the unique effective action 
on the arbitrary parameter $a$ of the term $g^{\mu\nu} g^{\al\be}$ 
of the configuration-space metric $G^{\mu\nu,\al\be}$ by the 
prescription that this metric is chosen as the metric contained
in the highest-derivative term of the bilinear form of the 
classical action in the minimal gauge. This choice is in 
consonance with the requirement that the metric 
in the space of the fields must be determined from the classical 
action, as proposed in the pioneer work~\cite{Vil-unicEA}.
We have shown that although this term changes under modified
parametrization of the quantum metric, the one-loop unique
effective action does not change. This confirms the consistency
of the mentioned additional requirement.

\section*{Acknowledgements}

The work of I.Sh. is partially supported by Conselho Nacional de
Desenvolvimento Cient\'{i}fico e Tecnol\'{o}gico - CNPq under the
grant 303635/2018-5.


\section*{Appendix: Generators of gauge transformations}
\label{ApA}

The gauge generators for the field $\phi_{\mu\nu}$ have been
evaluated in Ref.~\cite{JDG-QG} up to the zeroth order in $\ka$.
Nonetheless, we need the expansion up to the next order. The reason
is that the terms~\eqref{U1op} and~\eq{U2t} depend on the covariant
variational derivative of $R_{\mu\nu,\al}$ with respect to
$\phi_{\mu\nu}$, requiring the $O(\ka)$-approximation.

Consider the infinitesimal coordinate transformation
\beq
\label{CoorTrans}
x^\mu \, \longmapsto  \, x^{\prime\mu} \,=\, x^\mu + \xi^\mu.
\eeq
In the standard parametrization $\, g_{\mu\nu}^\prime$, the
generator reads
\beq
\label{ge-g}
R_{\mu\nu,\ga}^\prime (g^\prime)
\,=\, - (g_{\mu \ga}^\prime \na_\nu^\prime
+ g_{\nu \ga}^\prime \na_\mu^\prime ).
\eeq
The generators of gauge transformation for the quantum field
$\phi_{\mu\nu}$ can be obtained through a vector change of
coordinates in the space of the field representations,
\beq
\n{ge-trans}
R_{\mu\nu,\ga} (\phi)
\,=\,
\frac{ \pa  (\ka \phi_{\mu\nu})  }{ \pa  g^\prime_{\rho\si}}
\,R_{\rho\si,\ga}^\prime (g^\prime).
\eeq
By using Eqs.~\eq{gen-bfm}, \eq{ge-g} and \eq{ge-trans},
it is possible to show that
\beq
\label{ge-phi}
R_{\mu\nu,\ga} (\phi)
\,=\,
R_{\mu\nu,\ga}^{(0)} + \ka R_{\mu\nu,\ga}^{(1)} + O(\ka^2),
\eeq
where
\beq
\label{R0}
 R_{\mu\nu,\ga}^{(0)}
 \,=\,
 - \frac{1}{\ga_1} \left( g_{\mu\ga} \na_\nu
 + g_{\nu\ga} \na_\mu \right)
+ \frac{2 \ga_2}{\ga_1 (\ga_1 + D \ga_2)} \, g_{\mu\nu} \na_\ga
\eeq
and
\begin{equation}
\begin{split}
\label{R1}
 R_{\mu\nu,\ga}^{(1)} \,=& \, \,
  (r_1 - 1) \, ( \phi_{\mu \ga} \na_\nu + \phi_{\nu \ga} \na_\mu )
+ r_1 \, ( g_{\mu \ga} \phi_\nu^\la + g_{\nu \ga} \phi_\mu^\la ) \na_\la
+ r_2 \, g_{\mu\nu} \, \phi^\la_\ga \, \na_\la
\\
&
+ r_3 \, \phi_{\mu\nu} \na_\ga
-  (\na_\ga \phi_{\mu\nu})
+ r_4 \, \phi ( g_{\mu \ga} \na_\nu + g_{\nu \ga} \na_\mu )
+ r_5 \, g_{\mu\nu} \,  \phi \, \na_\ga ,
\end{split}
\end{equation}
with the coefficients
\beq
&&
r_1 = \frac{\ga_3}{\ga_1^3},
\qquad
r_2 = \frac{2 \ga^2_1\ga_2 - 4 (\ga_2 \ga_3 - \ga_1 \ga_4)}{\ga_1^2 (\ga_1 + D \ga_2)},
\qquad
r_3 = -\frac{2 (2 \ga_2 \ga_3 - \ga_1 \ga_5)}{\ga_1^2 (\ga_1+D \ga_2)},
\nn
\\
&&
r_4 = \frac{\ga_5 - \ga_1\ga_1}{\ga_1^2},
\qquad
r_5 = \frac{2 \ga_1\ga_2^2 + 4 \ga_2 (\ga_2 \ga_3 - \ga_1 \ga_4 )
- 2 \ga_2 \ga_5 (3 \ga_1 + D \ga_2)
+ 4 \ga_1^2 \ga_6}
{\ga_1^2 (\ga_1 + D \ga_2)}.
\mbox{\quad}
\nn
\eeq
The expressions~\eqref{R0} and~\eqref{R1} are sufficient for the one-loop
calculations reported in the main part of the paper.


\end{document}